\documentclass[preprint,preprintnumbers,aps,amsmath,amssymb,prb,superscriptaddress,floatfix]{revtex4-1}

\usepackage{graphicx}
\usepackage{dcolumn}
\usepackage{bm}
\usepackage[T1]{fontenc}
\usepackage[utf8]{inputenc}
\usepackage{physics}
\usepackage{color}
\usepackage{float}
\usepackage{natbib}
\usepackage{hyperref}
\usepackage{cleveref}

\newcommand{\vsrc}{$V_\mathrm{Src}$}

\newcommand{\vg}{$V_\mathrm{Gate}$}
\newcommand{\mudot}{$\mu_\mathrm{Dot}$}

\renewcommand{\figurename}{\textbf{Fig}}

\begin{document}

\title{Quantum-Dot Assisted Spectroscopy of Degeneracy-Lifted Landau Levels in Graphene}

\author{I. Keren}
\affiliation{Racah Institute of Physics, The Hebrew University, Jerusalem 91904, Israel}
\author{T. Dvir}
\affiliation{Racah Institute of Physics, The Hebrew University, Jerusalem 91904, Israel}
\author{A. Zalic}
\affiliation{Racah Institute of Physics, The Hebrew University, Jerusalem 91904, Israel}
\author{A. Iluz}
\affiliation{Racah Institute of Physics, The Hebrew University, Jerusalem 91904, Israel}
\author{D. LeBoeuf}
\affiliation{LNCMI, Centre National de la Recherche Scientifique, EMFL, Université Grenoble Alpes, INSA Toulouse, Université Toulouse Paul Sabatier, Grenoble, France}
\author{K. Watanabe}
\affiliation{National Institute for Material Science, 1-1 Namiki, Tsukaba 305-0044, Japan}
\author{T. Taniguchi}
\affiliation{National Institute for Material Science, 1-1 Namiki, Tsukaba 305-0044, Japan}
\author{H. Steinberg}
\affiliation{Racah Institute of Physics, The Hebrew University, Jerusalem 91904, Israel}

\date{\today}
Corresponding author: hadar@phys.huji.ac.il

\begin{abstract}

\textbf{Energy spectroscopy of strongly interacting phases requires probes which minimize screening while retaining spectral resolution and local sensitivity.  Here we demonstrate that such probes can be realized using atomic sized quantum dots bound to defects in hexagonal Boron Nitride tunnel barriers, placed at nanometric distance from graphene. 
With dot energies capacitively tuned by a planar graphite electrode, dot-assisted tunneling becomes highly sensitive to the graphene excitation spectrum.
The spectra track the onset of degeneracy lifting with magnetic field at the ground state, and at unoccupied exited states, revealing symmetry-broken gaps which develop steeply with magnetic field - corresponding to Land\'e $g$ factors as high as 160. 
Measured up to $B = 33$ T, spectra exhibit a primary energy split between spin-polarized excited states, and a secondary spin-dependent valley-split. Our results show that defect dots probe the spectra while minimizing local screening, and are thus exceptionally sensitive to interacting states.}

\end{abstract}

\maketitle

\section*{Introduction}

The use of electron tunneling as a spectroscopic probe for condensed matter systems was first demonstrated by Giaever~\cite{Giaever_1960}, who applied an oxide tunnel barrier to map the gap in the excitation spectrum of superconductors. Tunneling measurements involve a source electrode which couples to a sample system through a barrier - with the sample density of states (DOS) encoded in the differential change in the tunnel current at a finite source bias~\cite{Wolf_book}. Tunneling became a generic probe, able to address a broad range of conducting samples, following the introduction of the scanning tunneling microscope (STM)~\cite{Binnig_1982}.

Certain types of samples, however, challenge the existing tunneling methodology. For samples with low DOS, applied bias voltage leads to local charging effects, enhanced by differences in the work functions of the probe and sample. In graphene, for example, voltage applied to a local tunnel probe changes the potential landscape~\cite{Jung2011,Wang_Crommie_2012,Zhao_WGM_STM,Brar_2011}. 
At finite magnetic fields where graphene DOS becomes sharply peaked at Landau level energies $E_N = \pm v_\mathrm{F}\sqrt{2\hbar e|N|B}$~\cite{Zhang2005,Luican2011} ($N$ being level index), deformation of the potential isolates local regions due to incompressible strips~\cite{Wang_Crommie_2012}. This is further complicated by the need to reach elevated Landau levels, requiring bias voltages reaching well over 100 mV.

An ideal probe would be local in its physical extent, minimize screening of local interactions and at the same time retain a parallel geometry to avoid non-homogeneous charging. 
In addition, it should sustain a high bias without deforming local potentials. 
Here we show that these seemingly contradicting requirements are fulfilled by resonant tunneling through quantum dots (QDs) bound to atomic defects within van der Waals tunnel barriers~\cite{Greenaway2018, Dvir_PRL_2019}. Graphene-based tunnel junctions are highly parallel~\cite{Mishchenko_2014}, and dot energies are capacitively tuned by the planar electrode, thus avoiding local charging effects due to applied bias. 
Barrier-embedded dots are nanometric - both in their physical dimensions, and in their proximity to the sample layer. Thus, they are sensitive to regions few nm large.
Finally, defect dots lack any degree of freedom for charge rearrangement, they do not screen local interactions, and are hence less invasive than metallic probes. 

When QDs couple weakly to the source and drain electrodes, they permit resonant charge transport through sharply peaked energy levels. In this regime, sample DOS is probed by the current through the dot, rather than differential current - avoiding large DC contributions.
QDs are utilized as local thermometers~\cite{Venkatachalam_2012,Maradan_2014}, where energy distribution is tracked using the energy-selective nature of injection and ejection of carriers through the QD. 
Alternatively, by electrostatic tuning of the QD level, sequential tunneling through the QD singles out the sample DOS in resonance with the dot - as seen in the gate-tunable dots used to probe the spectra of superconductor-proximitized nanowires~\cite{Deng_2016,Junger_2019}.

We demonstrate the utility of QD-assisted spectroscopy in a study of the graphene excitation spectrum. In the graphene quantum Hall regime, Landau levels are fourfold degenerate.
This SU(4) symmetry allows for several distinct paths of symmetry breaking~\cite{Alicea_2006,Goerbig2006,Yang_2006,Nomura_2006,Kharitonov_2012,Fuchs_2007} driven by magnetic field, causing the emergence of ordered ground states which may break either spin or valley degeneracy. 
The order in which these symmetries should break is subject to debate: While the spin degeneracy is broken by the Zeeman effect, the breaking of valley degeneracy via magnetic field is not as straight-forward. It appears to depend on sample-specific properties such as disorder and the effect of interactions~\cite{Young2012}.  
Specifically, $N=0$ and $N \neq 0$ Landau levels differ in  wavefunction localization, causing a difference in the energy splitting due to lifting of the valley degeneracy. The magnitudes of the Zeeman effect and the short range interactions compete, resulting in different hierarchies between the spin and valley degeneracy lifting. 

So far, existing experiments sensitive to degeneracy lifting effects were provided by probes sensitive to the ground state. These include transport~\cite{Young2012,PhysRevLett.99.106802,PhysRevLett.96.136806} and STM measured near the Fermi energy~\cite{Song2010,Li_STM_2019}. 
The nature of degeneracy lifting in excited states remains an open question: A two-fold splitting of the filled $N=0$ level has been observed by STM~\cite{Song2010}, and it is indeed clear that excited energy levels should retain the Zeeman splitting. In excited state spectroscopy, carriers are injected into non-populated levels, or ejected from fully populated levels. In this scenario, any deviation from single-particle Zeeman splitting would indicate that energy levels are affected by inter-Landau-level interactions.  A non-trivial role of interactions will be manifest in two ways. First, any enhancement of the Land\'e $g$ factor from the non-interacting value, and second, the appearance of valley splitting in full or empty Landau levels.

\section*{Results}
\subsection*{Defect-Assisted Tunneling at Zero Magnetic Field}
In this work we report measurements of defect assisted transport between graphene and graphite separated by a hBN barrier. The barrier-defect energy is tunable by an electric field which originates from a top-gate and penetrates through the graphene layer. 
We carry out measurements up to magnetic fields of $B = 33$ T, and find an intricate pattern of lifting of both valley and spin degeneracies upon injection of carriers to the $N=0$ and $N=1$ excited Landau levels. The spectral splitting is dominated by a strongly enhanced Zeeman term, and valley-split energies are found to exhibit spin-valley coupling. 

\begin{figure*}[h]

\center
    
\includegraphics[width=\linewidth]{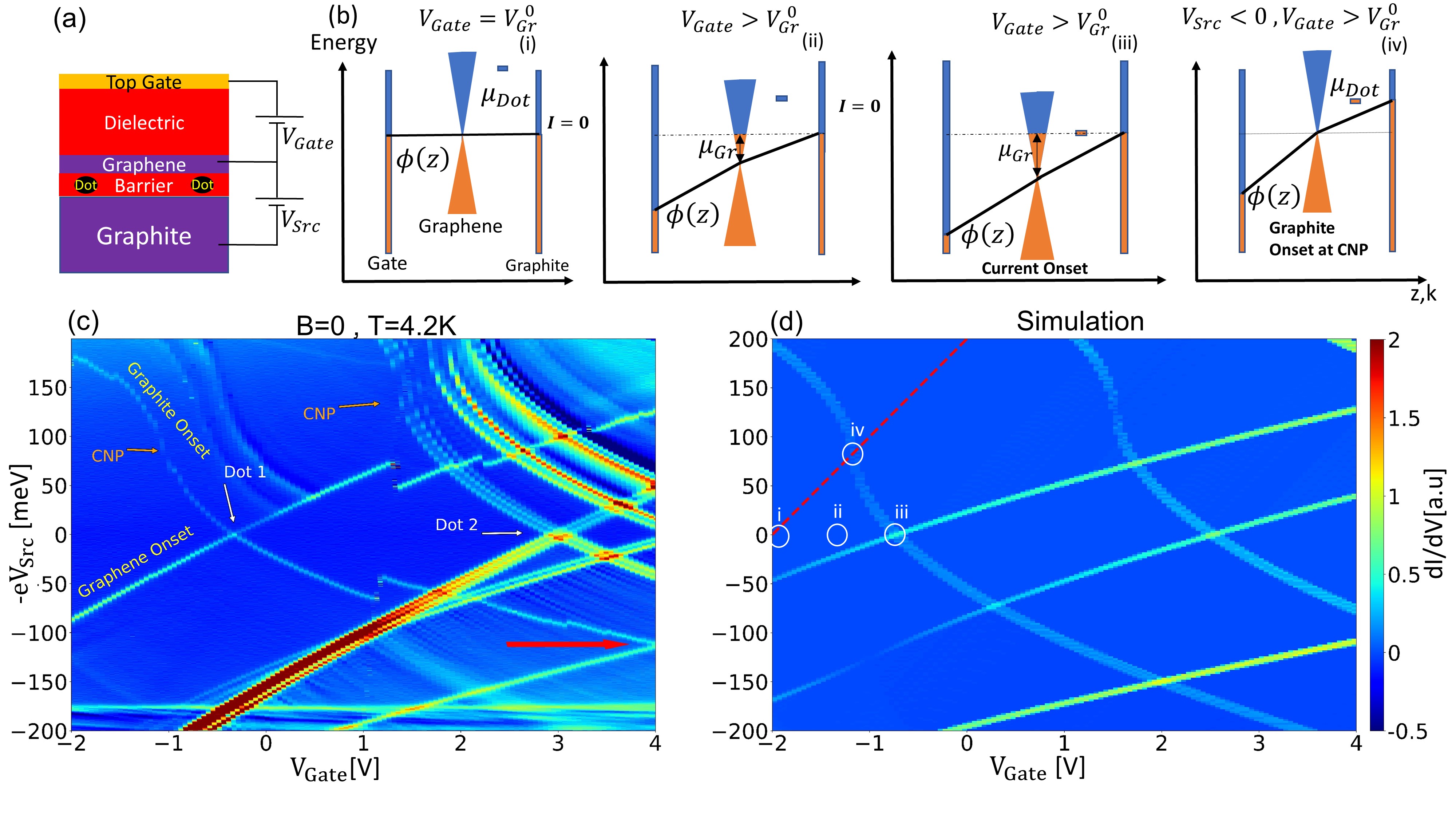}
        
\caption{ Tunneling through quantum dots (\textbf{a})  A schematic illustration of the device. The quantum dots are present within the hBN layer between graphite and graphene. (\textbf{b}) The energy diagram illustrating the change of the electrostatic potential $\phi(z)$ upon application of $V_\mathrm{Gate}$ and $V_\mathrm{Src}$ and its effect on the dot potential \mudot. The horizontal axis marks both $k$ (momentum of the graphene dispersion) and $z$ (the position coordinate).  (\textbf{c})  $\mathrm{d}I / \mathrm{d} V$ vs. $-\mathrm{e}V_\mathrm{Src}$ and \vg\ at $T = 4.2$ K. Transport signatures of distinct quantum dots, `Dot 1' and `Dot 2' are marked. CNP is marked by an arrow for both dots. Graphite (source) onset and graphene (drain) onset lines are labeled for Dot 1 while the red arrow signals the charging energy of Dot 1. (\textbf{d})  Simulated $\mathrm{d}I / \mathrm{d} V$ for the capacitive model described in the text, assuming two dots where $\mu_0 = 40$ meV  for Dot 1 and $\mu_0 = 67$ meV for Dot 2. The positions corresponding to diagrams (i)-(iv) in panel (b) are marked and so is an equal density line, corresponding to $n_\mathrm{Gr} = 0 $ near Dot 1 (red dashed line). }

\label{fig:Zero_Field}
        
\end{figure*}

Defects are regularly found in exfoliated materials, and their signatures have been observed via photoluminescence in transition metal dichalcogenides (TMD)~\cite{Chakraborty2015,He2015} and hexagonal Boron Nitride (hBN)~\cite{doi:10.1021/acsnano.6b03602} layers. 
Coupling defect-dots to source and drain electrodes entails placing an insulating layer between two conductors $-$ the same geometry used for tunnel junctions stacked using the vdW transfer technique~\cite{Dvir2018,Dvir2018b} (Figure \ref{fig:Zero_Field}(a)). 
This results in single charging behavior characteristic of quantum dots, as seen both in hBN and TMDs~ \cite{Chandni2015,Papadopoulos2019,Greenaway2018,Khanin2019}.
The dimension of a dot embedded within barriers depends on the type of defect and dielectric properties of the medium, and can range from the atomic size to a few nm \cite{Papadopoulos2019,Hong2015}. In addition, being embedded in a few layer insulator, barrier defects reside at nanometer proximity to both source and drain.

We report measurements taken on a device fabricated using the standard polycarbonate (PC) pickup method, schematically depicted in Fig.~\ref{fig:Zero_Field}(a) (See more details in Supplementary Note 5). The bottom (source) electrode is a graphite flake onto which a tunnel barrier hBN flake of thickness $d_\mathrm{Src} = 2$ nm (5 layers) is deposited. Graphene is picked up and placed on top of the barrier, capped by a second 20 nm hBN flake.  Ti-Au electrodes are deposited on the graphene and graphite, respectively, and a top-gate is patterned over the top hBN. Upon application of bias voltage \vsrc\ between the graphite and graphene flakes, the differential conductance $\mathrm{d}I / \mathrm{d} V$ measured at $T = 4.2$ K is dominated by several sharp features, which also depend on the gate voltage \vg. \vsrc\ and \vg\ both charge the graphene layer, adding a global charge density $\Delta n$, such that $n(x)=n_{0}(x)+
\Delta n(x)$ where $n(x)$ is the local density and $n_{0}(x)$ is the local density at $V_\mathrm{Gate}=V_\mathrm{Src}=0$. 
$\Delta n$ is calculated using the capacitive coupling of graphene to the source and gate.

A plot of $\mathrm{d}I / \mathrm{d} V$ vs. $-\mathrm{e}V_\mathrm{Src}$ and $V_\mathrm{Gate}$ is presented in Fig.~\ref{fig:Zero_Field}(c). It exhibits sharp features reminiscent of Coulomb-blockade diamonds, interpreted as the onset of resonant tunneling conductance through quantum dots embedded within the hBN barrier. It is immediately evident that the slopes of the differential conductance features in Fig. \ref{fig:Zero_Field}(c) are not constant. Such slopes are determined by the ratios of respective dot capacitances to the source, drain and gate electrodes~\cite{Greenaway2018}. The gate, however, is separated from the dot by the graphene layer, which screens its electric field. As this screening varies with the graphene density, the effective dot-gate capacitance varies as well. Field penetration through graphene is explained in the energy diagrams in Fig.~\ref{fig:Zero_Field}(b), where we schematically plot the evolution of graphene chemical potential $\mu_\mathrm{Gr}$, electrostatic potential $\phi(z)$, and dot chemical potential $\mu_\mathrm{Dot}$ with respect to applied \vg\ and \vsrc. (Here $z$ is the spatial coordinate perpendicular to graphene surface, with $z=0$ the position of the graphene layer. $\phi$ and $\mu_\mathrm{Gr}$ are defined as positive for electrons). 

\subsection*{Capacitive model for defect-assisted tunneling}
We take as a starting condition the neutrality point where graphene density is $n_\mathrm{Gr} = 0$, \vsrc\ = 0. At this condition we define $V_\mathrm{Gate} = V_\mathrm{Gr}^0$ (diagram (i)), where $V_\mathrm{Gr}^0$ is the voltage required to offset any background density in the graphene.
At neutrality, $\phi(z) = 0$ everywhere, and the dot energy is $\mu_0$ (we note that $\mu_0$ could vary depending on the type of defect \cite{Greenaway2018}).  The gate voltage affects the potential map by setting $\phi(-d_\mathrm{Gate}) = -\mathrm{e}V_\mathrm{Gate}$, $d_\mathrm{Gate}$ being the thickness of the gate dielectric, and $\mathrm{e}$ the absolute value of the electron charge. Applying positive gate voltage $V_\mathrm{Gate} > 0$ negatively charges the graphene layer. Throughout the measurement, graphene electrochemical potential $\mathrm{e}\mathrm{V_{Gr}}$ is kept at ground: $-\mathrm{e}V_\mathrm{Gr}=\mathrm{e}\phi(0)+\mu_\mathrm{Gr} = 0$, and hence the accumulation of negative charge results in $\mu_\mathrm{Gr}>0$ and a downward shift in $\phi(0)$. 

Changing the graphene chemical potential $\mu_\mathrm{Gr}$ modifies the dot energy, as is seen in diagrams (ii,iii). Here, an electric field $E = -\mathrm{d}\phi/\mathrm{d}z$ penetrates to the gap between graphene and the source electrode at position $z = d_\mathrm{Src}$. The dot, residing at position $z = d_\mathrm{Dot}$ will change it chemical potential to  $\mu_\mathrm{Dot}=\mathrm{e}\phi(d_\mathrm{Dot})+\mu_0$. Current will flow through the dot only if $0\leq\mu_\mathrm{Dot}\leq-\mathrm{e}V_\mathrm{Src}$ or $0\geq\mu_\mathrm{Dot}\geq-\mathrm{e}V_\mathrm{Src}$. 
Conduction onset at zero-bias will take place when the dot is resonant with both graphite source and graphene drain (diagram (iii)). In diagram (iv) we plot the condition of current onset at finite \vsrc\ where $-\mathrm{e}V_\mathrm{Src} = \mu_\mathrm{Dot}$. Here $V_\mathrm{Src} < 0$ is applied concomitantly with $V_\mathrm{Gate} > 0$ to keep graphene density constant. 

The source onset condition, where the dot potential is resonant with the source, is sensitive to changes in graphene chemical potential $\mu_\mathrm{Gr}$. This is seen by writing the dot potential as (Supplementary Note 1)
\begin{equation}\label{eq_mu_dot}
\mu_\mathrm{Dot}=\mu_0+\mu_\mathrm{Gr}\left(\frac{d_\mathrm{Dot}}{d_\mathrm{Src}}-1\right)-\mathrm{e}V_\mathrm{Src}\frac{d_\mathrm{Dot}}{d_\mathrm{Src}}
\end{equation}
Using the source-onset condition, we find that $V_\mathrm{Src}$ traces $\mu_\mathrm{Gr}$ up to a constant along the source onset line:
\begin{equation}\label{eq:Vsrc}
    V_\mathrm{Src}=\frac{\mu_\mathrm{Gr}}{\mathrm{e}}+\frac{\mu_0}{\mathrm{e}}\frac{d_\mathrm{Src}}{d_\mathrm{Dot}-d_\mathrm{Src}}
\end{equation}
As we show in Supplementary Equation~11, this condition can be used to extract compressibility from the slope of the source onset line. As expected for single-layer graphene, the source-onset line $V_\mathrm{Src}(V_\mathrm{Gate})$ traces a square-root dependence.

In Fig.~\ref{fig:Zero_Field}(c) we identify the transport signatures of two distinct dots (Dot 1 and Dot 2), which likely reside in different regions and conduct in parallel (upon a 2nd cooldown, Dot 1 has shifted in gate voltage with respect to Dot 2).
From Supplementary Equation 11, the maximal slope (absolute value), marked in the figure, is found where the graphene layer near each dot reaches the charge-neutrality point (CNP). 
Dot 1 has lower $\mu_0$, and is accessible at lower density. Its charging energy, estimated from the crossing marked by a red arrow in the figure, is $U \sim 110$ meV. We note that this is a higher value than found in similar studies \cite{Papadopoulos2019,Hakonen}. Dot 2 appears at a higher density. It exhibits an energy splitting even at $B = 0$ T (The origin of which is not presently understood). 

Transport through the barrier dots is simulated by calculating a capacitive model of double-gated graphene \cite{sanchez} with a fermi velocity of $v_\mathrm{F}=1.1\times 10^6$ $\mathrm{m \, s^{-1}}$ resulting in the differential conductance map shown in Fig.~\ref{fig:Zero_Field}(d). In this model, capacitive charging is induced by the source and gate voltages following: 
\begin{equation}\label{eq: top and bottom charge}
    -en_\mathrm{Gate,Src}=C_\mathrm{Gate,Src}(V_\mathrm{Gate,Src}-\frac{\mu_\mathrm{Gr}}{\mathrm{e}})
\end{equation} Where $n_\mathrm{Gate,Src}$ are the charge densities accumulated on the gate and source electrodes respectively, $C_\mathrm{Gate,Src}$ are the corresponding capacitances. 
The total charge is fixed to an initial $n_0$. 
\begin{equation}\label{eq: charge neutrality}
    n_\mathrm{Gate}+n_\mathrm{Src}+n_\mathrm{Gr}=n_0
\end{equation} 
and $n_\mathrm{Gr}$ is related to $\mu_\mathrm{Gr}$ by an integral on the DOS, $\rho(E)$:
\begin{equation}\label{Eq_mug}
    n_\mathrm{Gr}=\int_{0}^{\mu_\mathrm{Gr}}\rho(E)dE
\end{equation}

Together, \Crefrange{eq: top and bottom charge}{Eq_mug} yield an integral equation for $\mu_\mathrm{Gr}$ : 
\begin{equation}\label{eq_simulation}
\int_{0}^{\mu_\mathrm{Gr}}\rho(E)dE = n_0 + \frac{C_\mathrm{Gate}V_\mathrm{Gate}+C_\mathrm{Src}V_\mathrm{Src}}{\mathrm{e}}-\frac{\mu_\mathrm{Gr}}{\mathrm{e}^2}(C_\mathrm{Gate}+C_\mathrm{Src})
\end{equation}

By numerically solving \Cref{eq_simulation} while using the known DOS of graphene $\rho = \frac{2}{\pi v_\mathrm{F}^2 \hbar^2} |E|$, we find $\mu_\mathrm{Gr}(V_\mathrm{Gate},V_\mathrm{Src})$. 
Extracting $\mu_\mathrm{Dot}$ using \Cref{eq_mu_dot}, we obtain a differential conductance map for the contribution of each dot. From the simulation we extract $\mu_0 = 40$ meV for Dot 1 and $\mu_0 = 67$ meV  for Dot 2. For both dots $d_\mathrm{Dot} = 1$ nm. Although we have no information about the chemical identity of the defects, the capacitive model places both of them at the center of the 5 layers. 

\subsection*{Landau Level Spectroscopy}
\begin{figure*}[!htb]
\includegraphics[width=\linewidth]{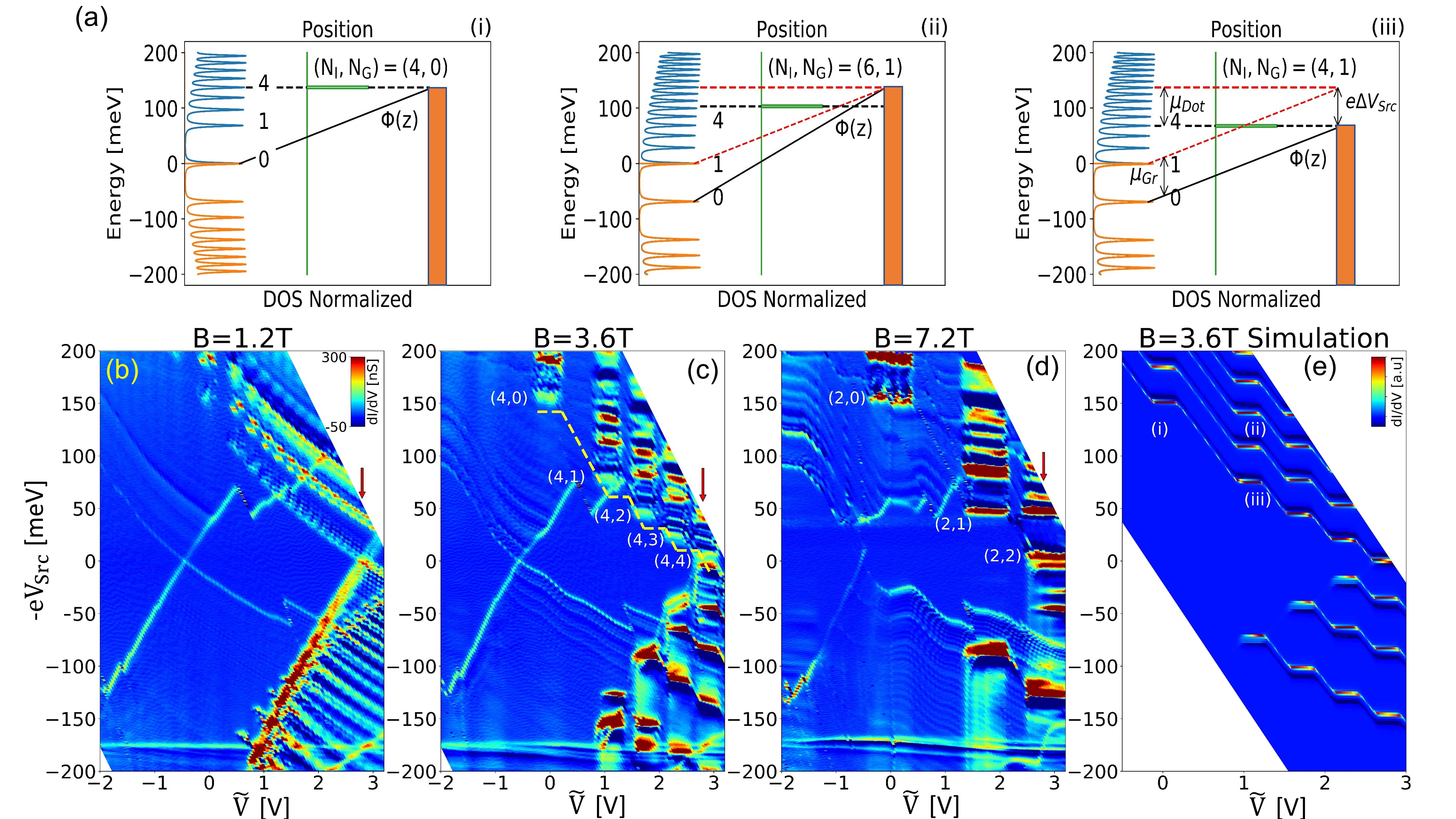}  
        
\caption{ Tunneling through a quantum dot at finite magnetic field (\textbf{a})  Energy diagrams describing dot energy (black dashed line) with respect to graphene spectrum (orange - occupied levels, blue - unoccupied levels). 
Panels (i)-(iii) correspond to the transition from $(N_\mathrm{I},N_\mathrm{G})=(4,0)$ to $(4,1)$.
The transition from (i) to (iii) is decomposed into two steps: Constant bias (i,ii) where $(N_\mathrm{I},N_\mathrm{G})$ changes from $(4,0)$ to $(6,1)$, and constant density (ii,iii) where the injection level is recovered ($(N_I,N_G)=(4,1)$). 
The red dashed line reflects $\phi(z)$ and $\mu_\mathrm{Dot}$ at (i). 
Variation in $\mathrm{e}V_\mathrm{Src}$ between (i) and (iii) equals the change in $\mu_\mathrm{Gr}$.
(\textbf{b-d})  $\mathrm{d}I / \mathrm{d} V$ maps at $B=1.2$ T, $B=3.6$ T and $B=7.2$ T. The dashed line in (c) marks a simulated trajectory where injection level remains constant while varying the ground state. The horizontal axis shows $\tilde{V}$, a linear combination of \vsrc\ and \vg\ such that the density is constant along vertical lines. Red arrows mark $\tilde{V}=2.8$ V which corresponds to $n_\mathrm{Gr}=3.1\times 10^{12}$ $\mathrm{cm^{-2}}$, where Fig.~\ref{Fig_Lfan_fixed_n}(a) is measured. (\textbf{e}) Simulated $\mathrm{d}I / \mathrm{d} V$ map at $B=3.6$ T shows the same step structure as the data. The steps corresponding to illustrations (i)-(iii) in (a) are marked.}
   
\label{Figure_Low_Field}
    
\end{figure*}

We now turn to the effect of perpendicular magnetic field $B$ on the transport through the quantum dot. In Fig.~\ref{Figure_Low_Field}(b-d), we plot $\mathrm{d}I / \mathrm{d} V$ maps while applying magnetic fields of $B = 1.2$ T, $3.6$ T, and $7.2$ T, respectively. The horizontal axis shows $\tilde{V}=V_\mathrm{Gate}+\frac{C_\mathrm{Src}}{C_\mathrm{Gate}}V_\mathrm{Src}$ - a linear combination chosen such that graphene equal density lines are vertical. 
The same data, presented vs. $V_\mathrm{Gate}$, appear in Supplementary Figure 1.
At the quantum Hall regime the onset of the dot transport attains a step-like structure in the ($\tilde{V}$,$-\mathrm{e}V_\mathrm{Src}$) plane, characterized by flat conductance features whose width along the $\tilde{V}$ axis increases due to increasing Landau level degeneracy. To further elucidate the structure of these features, we focus on Dot 2, where steps are sharpest. 

The origin of the step structure is in the Landau level DOS, as seen in Fig.~\ref{Figure_Low_Field}(a) which depicts the energy diagram at a finite magnetic field. The DOS in the quantum Hall regime consists of discrete energy levels broadened due to disorder. In this regime, the onset of electron injection into graphene takes place when $\mu_\mathrm{Dot} < -\mathrm{e}V_\mathrm{Src}$ and at the same time the dot is resonant with an unpopulated Landau level:$-\mu_\mathrm{Gr}+E_N = \mu_\mathrm{Dot}$. 
A similar condition holds for electron ejection at the opposite bias.
Panel (i) depicts such a resonant condition with electrons injected into Landau level $N=4$. At the same time, the graphene Fermi energy resides within the $N=0$ Landau level. Since the measured spectrum depends both on the graphene ground state, and on the injection state, which are generally not the same, we designate each spectral feature by the respective pair of ground state and injection indices $(N_\mathrm{I},N_\mathrm{G})$. $N_\mathrm{I} > N_\mathrm{G}$ ($N_\mathrm{I} < N_\mathrm{G}$) for injection (ejection) of electrons.

The $N_\mathrm{G}=0$ feature of Dot 2 is found at ($\tilde{V} = 0$, $-\mathrm{e}V_\mathrm{Src}=155$ meV) at all magnetic fields.
The dashed line in Fig.~\ref{Figure_Low_Field}(c) is the source-onset line in this field. Along this line the dot injects carriers to the same excited state $N_\mathrm{I}$, while the graphene electron density increases such that $N_\mathrm{G}$ changes from $0$ to $4$. This trajectory can be traced until $V_\mathrm{Src}=0$, where the dot is resonant with the source and the drain, and electrons are injected into the ground state. At $V_\mathrm{Src}=0$ we have $N_\mathrm{I}=N_\mathrm{G}$, identifying $N_\mathrm{I} = 4$ for this entire trajectory.

The (4,0) feature thus corresponds to the compressible regime, where $\mu_\mathrm{Gr}$ resides within the $0^{\mathrm{th}}$ Landau level, the DOS is very large, and graphene perfectly screens the electric field applied by the gate. Once this level is filled, graphene enters the incompressible regime. Its carrier density can not change - allowing almost perfect field penetration. As a result, $\phi(z=0)$ shifts sharply downwards. The next compressible regime, corresponding to the ground state $N_\mathrm{G}=1$, appears at $-\mathrm{e}V_\mathrm{Src} = 78$ meV. 

Between (4,0) and (4,1), the graphene ground state changes while the dot is kept resonant with the same injection level. Using \Cref{eq_mu_dot} we 
show in the Supplementary Notes 2,3 that along such a trajectory $\Delta\mu_\mathrm{Gr} = \mathrm{e}\Delta V_\mathrm{Src}$, where $\Delta\mu_\mathrm{Gr}$ is the change in graphene chemical potential, and $\Delta V_\mathrm{Src}$ is the change in \vsrc\ required to keep the same Landau level resonant. 
Using this relation, we find that the difference between the $N_\mathrm{G} = 0,1$ plateaus is $77$ meV, in agreement with the parameters used for the fit in Fig.~\ref{fig:Zero_Field}(d). More generally, we can simulate the entire spectrum using the same model developed above (\Cref{eq_simulation}), while assuming a density of states consisting of Gaussian broadened Landau levels. 
From the simulation we extract the source onset line corresponding to the $N_\mathrm{I} = 4$ trace. Plotted as a yellow dashed line in Fig.~\ref{Figure_Low_Field}(c), this trace agrees very well with experimental data.

\subsection*{Degeneracy Lifting}
The correspondence between $V_\mathrm{Src}$ and $\mu_\mathrm{Gr}$ (\Cref{eq:Vsrc}) shows that the source-onset line is sensitive to changes in $\mu_\mathrm{Gr}$ and can hence be used as a compressibility probe (Supplementary Equation~11). In this sense, the barrier-defect functions as a very local single-electron transistor (SET)~\cite{RevModPhys.64.849,Yoo1997,Wei1997,Ilani2004}.
In what follows, we show that the dot also serves as a spectral probe, sensitive to excited state DOS. 
The dot is used as a spectrometer by keeping $n_\mathrm{Gr}$ constant and scanning the injection energy, as explained schematically in panels (ii,iii) of Fig.~\ref{Figure_Low_Field}(a). Here, $n_\mathrm{Gr}$ is kept constant by balancing \vg\ and \vsrc, the dot energy follows $\Delta\mu_\mathrm{Dot}=(d_\mathrm{Dot}/d_\mathrm{Src})\Delta V_\mathrm{Src}$. Thus, keeping the ground state fixed, the dot maps the spectra of different injection levels $N_\mathrm{I}$. 

\begin{figure*}[ht]
\includegraphics[width=\linewidth]{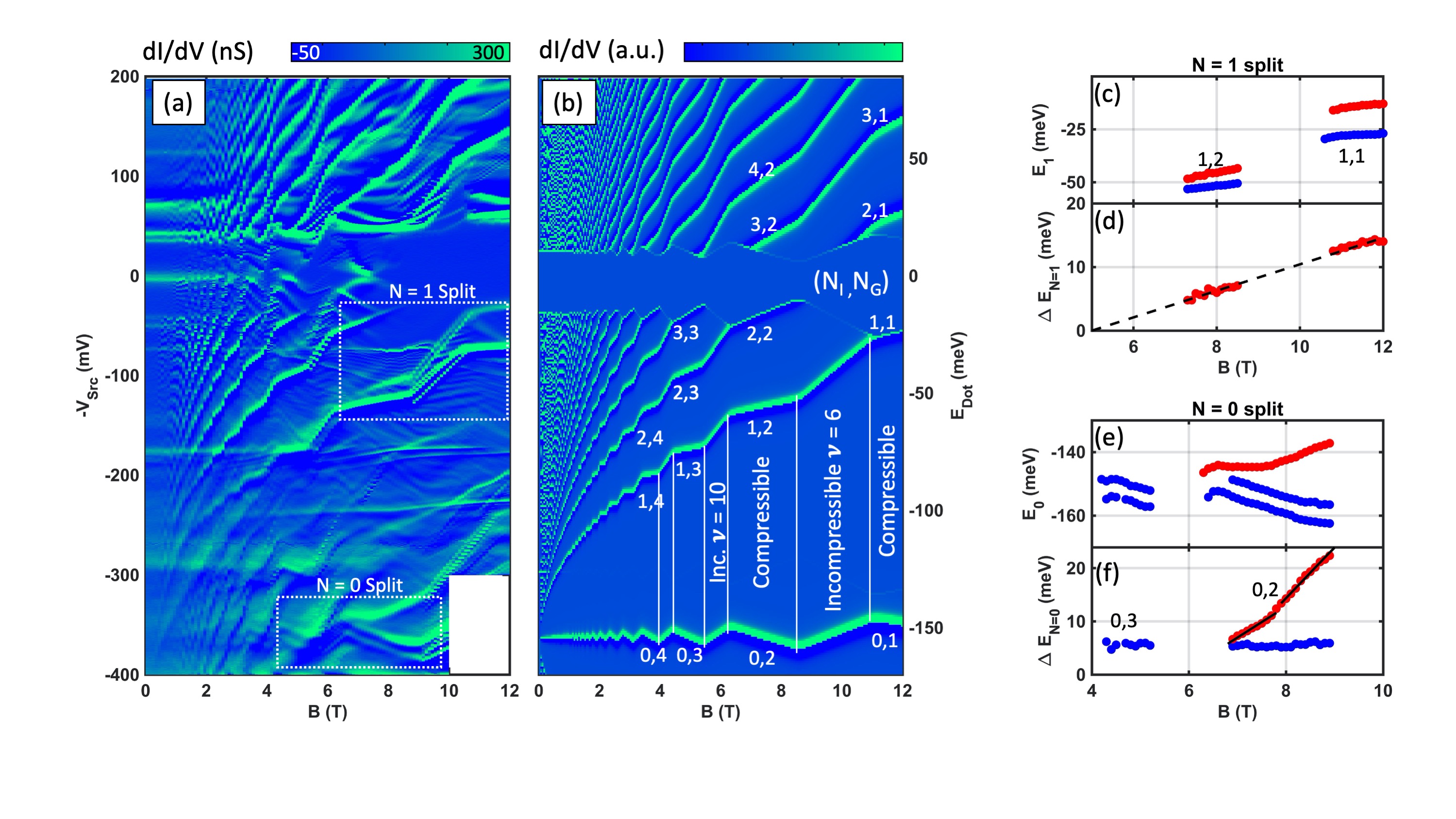}
\caption{Spectrum at fixed density. \textbf{(a)} $\mathrm{d}I / \mathrm{d} V$ vs. -\vsrc\ and $B$, graphene density fixed at $n_\mathrm{Gr} = 3.1\times 10^{12}$ $\mathrm{cm^{-2}}$. Dashed regions mark $N=0$ and $N=1$ spectral features which show evidence of degeneracy lifting. \textbf{(b)} Simulation modelling the data at (a). $\mu_\mathrm{Dot} = 95$ meV, $d_\mathrm{Dot} = 0.85$ nm. Spectral features are marked by index pair: Injection / ejection level $N_\mathrm{I}$ and ground state $N_\mathrm{G}$. Compressible and incompressible regimes are marked at the negative bias. (c) Energies of the split peaks of the $N=1$ feature. (d) Energy difference of the features in (c). Dashed line corresponds to $g=36$. (e) Energies of the split peaks of the $N=0$ feature. (f) Energy difference of the peaks in (e). Blue: Valley split. Red: Spin split. Overlay lines trace two different slopes, corresponding to $g=100$ and $g=160$ respectively.}
\label{Fig_Lfan_fixed_n}        
\end{figure*}

Barrier-dot-assisted spectroscopy is demonstrated in Fig.~\ref{Fig_Lfan_fixed_n}(a), where density is kept fixed at $n_\mathrm{Gr} = 3.1\times 10^{12}$ $\mathrm{cm^{-2}}$ near dot 2 (marked by arrows in Fig.~\ref{Figure_Low_Field}(b-d)). \vsrc\ is scanned from $-400$ to $+200$ mV, corresponding to dot energies $E_\mathrm{Dot} = -180$ to $90$ meV (axis on the right of panel (b)).
The spectrum is dominated by Landau levels whose energies follow the well known $\sqrt{B}$ dependence, with kinks appearing as the graphene ground state shifts between compressible and incompressbile regimes. The simulation (panel (b)), based on the same model used above, reproduces this spectral map with excellent fidelity. 
The sharply peaked dot DOS causes the spectra measured this way to be extremely stable, since the DC contribution from levels below the dot energy is strongly suppressed. Compared to spectra measured using STM\cite{Luican2011,Stroscio2012}, it is seen here that the dot-assisted tunneling produces clear spectra at high bias, with well-resolved Landau levels at energies well over 150 meV. As we show below, the use of this probe on high quality encapsulated graphene samples reveals energy splitting related to the SU(4) degeneracy lifting in excited states.

In Fig.~\ref{Fig_Lfan_fixed_n}(a), two regions are marked, where the observed spectrum deviates from the calculation appearing in (b). In these regions, which correspond to injection of holes to the $N_\mathrm{I}=1$ and $N_\mathrm{I}=0$ levels, the spectral features are split due to lifting of the fourfold spin-valley degeneracy. The split spectral features of the $N_\mathrm{I}=0$ state are extracted and plotted separately in panel (e). The spectrum consists of two peaks, which exhibit a separation of 6 meV visible at fields as low as $B = 4$ T. This split remains fixed all the way to 9 T - suggesting a spin-independent origin. 
Since in $N_\mathrm{G} = 0$ the valley and sub-lattice degrees of freedom are coupled, this split could be due to a local breaking of sub-lattice symmetry - e.g. due to substrate effects, which should be independent of magnetic field.

At 6 T, a third spectral feature becomes visible. This feature, marked in red in panel (e), opens a gap which broadens rapidly (panel (f)). The gap, which evolves linearly with magnetic field, reaches a value of 17 meV at 9 T.  
It exhibits a very high slope: Extracting the $g$ factor using $\Delta E_\mathrm{Z}=g\mu_{\mathrm{B}}B$ (where $\mu_{\mathrm{B}}$ is the Bohr magneton), the observed split corresponds to $g\approx 100$ for $7 < B < 8$ T, and $g\approx 160$ for $8 < B < 9$ T (Fig.~\ref{Fig_Lfan_fixed_n}(f)). 
We can compare these values to STM measurements. For single-layer graphene on SiC~\cite{Song2010}, the $N=0$ gap stands at 20 meV at 13 T. Another study, on tri-layer graphene, finds $g = 14$.~\cite{He2018} Here we find that in our device the gap is both larger, and develops faster in magnetic field. 
Another large splitting appears at the $N_\mathrm{I}=1$ feature. Extracting the split peak energies (Fig.~\ref{Fig_Lfan_fixed_n}(c)), we find an energy gap following a linear dependence with $g = 36$ (Fig.~\ref{Fig_Lfan_fixed_n}(d)).
 
 The origin of such large Zeeman splits is puzzling. In atomic defects $g\approx 2$,~\cite{Dvir_PRL_2019} so the split is hence likely to originate in the graphene layer. Quantum Hall states are known to exhibit strong enhancements of the $g$ factors, associated with exchange coupling~\cite{Abanin_2006, Sondhi_1993}. 
$g$ may vary with sample quality, increasing in cleaner samples where a larger number of carriers may be polarized~\cite{Young2012}. 
The exchange energy split, predicted in Ref.~\cite{Abanin_2006} is $\approx 40$ meV at 10 T - not far from the values we measure.
Here, both the $N_\mathrm{I}=1$ and $N_\mathrm{I}=0$ states exhibit a split feature which extrapolates to zero at $B = 5$ T, where $N_\mathrm{G} = 3$. We thus speculate that the observed splitting is related to an interplay between the state of the injected carriers, and the onset of a spontaneous polarization at the $N_\mathrm{G}=3$ state.
In this scenario, the excited holes injected at the $N_\mathrm{I}=0,1$ state undergo a strong exchange interaction with polarized spins in the ground state. This is plausible, since carriers at different Landau levels occupy the same spatial coordinates.
Alternatively, the barrier dot which is 1 nm away from the graphene layer may itself experience strong exchange coupling to the graphene layer.
Distinguishing between these two models requires calculations which are beyond the scope of the present work.

\begin{figure*}

    
\includegraphics[width=\linewidth]{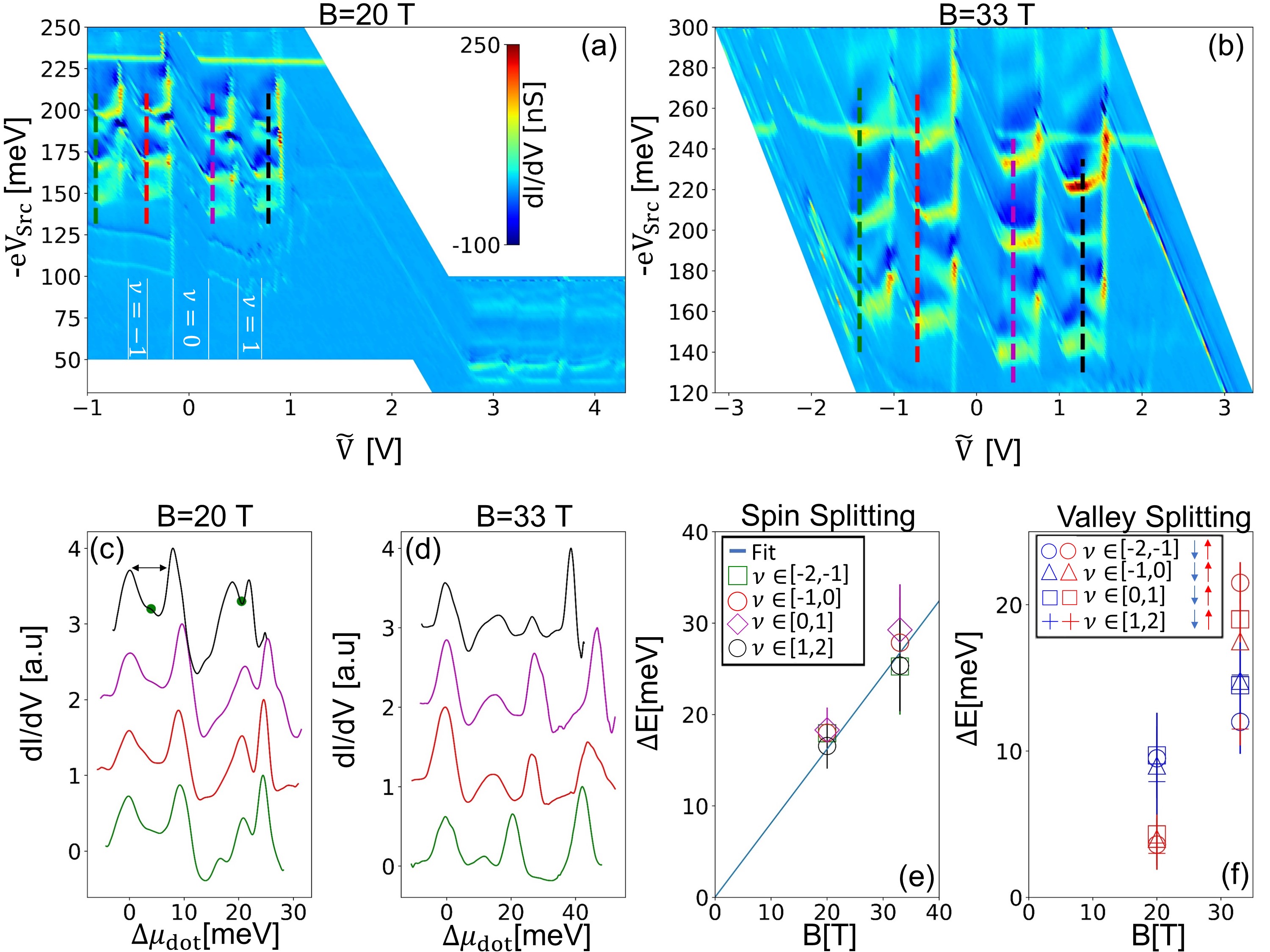}

\caption{Degeneracy Lifting at High Fields. \textbf{(a)} $(N_\mathrm{I} = 1,N_\mathrm{G} = 0)$ (upper left) and $(N_\mathrm{I} = 1,N_\mathrm{G} = 1)$ (lower right) at $B=20$ T. Splitting along the $\tilde{V}$ axis corresponds to degeneracy lifting of the ground state, while splitting along the $-\mathrm{e}V_\mathrm{Src}$ axis corresponds to splitting of the injection state. Regions of integer filling factor are marked. \textbf{(b)} $(N_\mathrm{I} = 1,N_\mathrm{G} = 0)$ at $B=33$ T. \textbf{(c,d)} $\mathrm{d}I / \mathrm{d} V$ vs. dot energy along the colored lines in panels a,b. The means of the low and high bias pairs of peaks are marked by green dots. The energy difference within each pair is marked by the double-headed arrow. \textbf{(e)} The energy difference between the mean of the low and high energy two peaks, for every colored line (fill factor) in panels a,b. Error bars correspond to half the peak width. The differences are plotted along a linear fit to the Zeeman energy splitting with a fit to $g = 14 \pm 1$. \textbf{(f)} The energy difference between same-spin peaks vs. $B$. Blue markers indicates the down-spin peaks, and red markers indicate the up-spin peaks.  }

\label{fig:High_Field}
\end{figure*}

Turning our attention to energy splitting where the $N=0$ level is tuned to the ground state, we notice that the feature marked by ($N_\mathrm{I}=2$,$N_\mathrm{G}=0$) in Fig.~\ref{Figure_Low_Field}(d) already shows incipient signatures of degeneracy lifting at $B=7.2$ T. 
Further increase of the magnetic field resolves the structure of the ($N_\mathrm{I},0)$ feature. The data, presented in Fig.~\ref{fig:High_Field}, were taken at the high magnetic field facility in Grenoble at temperature $T=1.2$ K. In panel (a) we plot the spectra measured at $B=20$ T. We find spectral features corresponding to $N_\mathrm{G}=0$ and $N_\mathrm{G}=1$. 
We focus on the $N_\mathrm{G}=0$ manifold, where the continuous feature found at lower fields has separated into an intricate pattern consisting of 16 distinct features. 
The origin of this structure is in the four-fold degeneracy lifting of both ground state $N_\mathrm{G}$ and injection state $N_\mathrm{I}$: As we've seen in Fig.~\ref{Fig_Lfan_fixed_n}, tuning $V_\mathrm{Src}$ maps the excited state spectrum at the injection level - in this case $N_\mathrm{I}=1$. 
At this high magnetic field, all four levels are distinguishable - as seen also at the fixed-density line-cuts, presented in panel (c). Along the horizontal ($\tilde{V}$) axis, the breakup into four features is a consequence of degeneracy lifting in the $N_\mathrm{G}=0$ ground state. The spectrum clearly exhibits narrow incompressible regions, corresponding to fill factors $\nu=-1,0,1$.

To understand which degeneracy drives the dominant splitting, we measure the energy difference between the mean of the two low energy peaks and the mean of the two high energy peaks of the spectra in Fig.~\ref{fig:High_Field}(c,d). The differences between these means (green dots in panel (c)), $\Delta E$, are plotted in Fig.~\ref{fig:High_Field}(e). For each of the available magnetic field data sets (B = 20, 33 T), we have four data points - corresponding to the different ground states $\pm\nu\in[1,2]$ and $\pm\nu\in[0,1]$. As seen in panel (e), at each magnetic field all four points are bunched closely together. A single fit for $\Delta E(B)$ can be used for all fill factors. The fit follows a straight line which extrapolates to zero, implying that Zeeman splitting is the leading degeneracy lifting term, in agreement with early transport measurements~\cite{PhysRevLett.96.136806,PhysRevLett.99.106802}. The slope  yields $g=14\pm1$. Splitting the $N=1$ level should be compared to transport results with a ground state fill factor $\nu=4$. We find a greater $g$ factor than those found before, where excitations at $\nu=-4$ yielded $g=7$.~\cite{Young2012} A number of causes could explain this difference. First, in Ref~\cite{Young2012}, the gap is measured via temperature dependence of a macroscopic sample, with $N=1$ being the ground state. Here, we measure the spectrum of an excited state, with $N=0$ being the ground state. Second, our measurement is local, and is hence less prone to averaging effects of disorder.

Having identified the leading split with the Zeeman term, we associate the two lower energy states with spin up, and the two higher energy states with spin down. In Fig.~\ref{fig:High_Field}(f) we plot the energy difference within each same-spin pair. It is evident that in all fill factors, the gap between spin-down states (blue markers) increases, but by a lesser amount with respect to the gap between spin up states (red markers).  
This result indicates that the valley splitting depends on the spin state, suggesting a coupling between the valley and spin degrees of freedom. 
Coupling between spin and valley degeneracy lifting has been discussed in graphene quantum dots localized by STM tips~\cite{Freitag_2016,Li_STM_2019,Luican_2014_STM}, where strong confinement lifts orbital degeneracy. 
Our system lacks a strong confining potential. Even if the dot is charged, the charging field should be screened at the compressible limit. 

Here we suggest that valley splitting could be indicative of the nature of the ground state. In the Anti-ferromagnetic (AF) or canted anti-ferromagnetic (CAF) ground states, for example, spin is coupled to the sub-lattice degree of freedom, which translates to the valley degree of freedom at $N=0$. In Fig.~\ref{fig:High_Field}, electrons are injected into the $N_\mathrm{I}=1$ level while $N_\mathrm{G}=0$. 
We speculate that for the single-particle state of the injected spin-polarized electron, the valley degree of freedom will dictate its relative overlap with the spins of the many-body ground state. 
While this calls for further calculation, it is likely the ferromagnetic (F) state can be ruled out when such splitting is observed.

\section*{Discussion}
Based on the compiled values of degeneracy lifted energy splitting (Table~\ref{table:g_compilation}) we can conclude that exchange interactions appear to play a major role in the spin-split state in high quality samples. The spin-split features measured in Fig.~\ref{Fig_Lfan_fixed_n} develop within a very small window in magnetic field - suggesting some pre-condition for their onset. In addition, we find that the ground state plays an important role in determining the excited state spectrum. The $N_\mathrm{I}=1$ feature develops differently when the ground state is $N_\mathrm{G}=0$ ($g=14$) and when $N_\mathrm{G}=1,2$ ($g=36$). This could be possible if, indeed, the splitting is governed by exchange interactions - since the injected / ejected carrier will experience the strongest interactions with carriers in the ground state. Finally, we find that a non-trivial interplay exists between spin and valley splitting. 
\begin{table}
\begin{center}
 \begin{tabular}{||c c c c c||} 
 \hline
 $N_\mathrm{G}$ & $N_\mathrm{I}$ & $g$ & $B(T)$ &\\ [0.5ex] 
 \hline\hline
 2,3 & 0 & 100-160 & 4-9 & Fig.~\ref{Fig_Lfan_fixed_n} \\ 
 \hline
 1,2 & 1 & 36 & 6-12 & Fig.~\ref{Fig_Lfan_fixed_n}\\
 \hline
 0 & 1 & 14 & 20-33 & Fig.~\ref{fig:High_Field}\\
 \hline
\end{tabular}
\end{center}
\caption{Compilation of $g$ values measured for degeneracy lifted states.}
\label{table:g_compilation}
\end{table}

At high magnetic fields, we also notice that the barrier dot becomes sensitive to the transition between compressible and incompressible regimes. As seen in  Fig.~\ref{fig:High_Field}(a,b), upon filling each Landau level the dot energy shifts vertically up before turning down again. 
We suggest that the sharp evolution of the density-dependent dot spectrum is a consequence of dot sensitivity to local disorder potential, which forms electrically floating regions in the graphene layer. 
Since the observed feature appears close to the full Landau level limit, the floating compressible island should be hole-like, and is coupled to a local potential maximum. Interestingly, such maximum could be induced by the negative charge of the dot itself~\cite{Luican_2014_STM,Wong_Defects_2015}. Confirming this, however, will require further study with additional samples.

Our results demonstrate the efficacy of barrier-dot-assisted spectroscopy as a probe for graphene in strongly-interacting quantum Hall states. The wide values of energy splitting away from the ground state suggest that strong exchange interaction has to be considered between excited carriers and polarized ground states. 
The observation of such features indicates that the barrier dot, as a probe, retains fragile many-body states. This could be the consequence of minimal screening, planar geometry, or both. 
As barrier dots may also serve as local sensors for the chemical potential, they thus merge the capabilities of local SETs~\cite{RevModPhys.64.849,Yoo1997,Wei1997,Ilani2004} with probes which retain planar geometry~\cite{Dial_2007}. 
Their size, positioning and non-invasiveness thus make barrier dots potentially useful probes for other interacting systems.

\section*{Methods}

\subsection*{Device Fabrication}
The vdW tunnel junction was fabricated using the vdW transfer technique. The graphite flake was exfoliated on a SiO$_2$ substrate. The hBN barrier and graphene flakes were transferred on top of the graphite flake, respectively. On top of the graphene flake, another bulk hBN flake was transferred, with the purpose of acting as a gate dielectric. Ti/Au contacts were fabricated using standard electron beam lithography techniques. Contact evaporation was executed at high vacuum and at $-5^\circ$ C.

\section*{Acknowledgements}
The authors are thankful for  discussions with S. Ilani, D.Orgad, A. Yacoby, P. Jarillo-Herrero, E. Rossi and E. Andrei.
M. Aprili, C. H. L. Quay and M. Kuzmenovic assisted with high magnetic field measurements. Device fabrication and characterization were carried out at the Harvey M. Krueger Family Center for Nanoscience and Nanotechnology. 
Part of this work was performed at the LNCMI, a member of the European Magnetic Field Laboratory.
Work was supported by ERC-2014-STG Grant No. 637298 and ISF Quantum Initiative grant No. 994/19.
A.Z. and T. D. Are supported by an Azrieli Fellowship.
K.W. and T.T. acknowledge support from the Elemental Strategy Initiative conducted by the MEXT, Japan and and the CREST (JPMJCR15F3), JST.

\section*{Author Contributions}
I. K., T. D. and A. I. fabricated the device. Measurements were carried out by I. K. and A. Z.. Data were analysed by I. K.. High Field measurements were done by D. L.. K. W. and T. T. provided the hBN. Manuscript was drafted by I. K. and H. S.. H. S. supervised the project. 

\section{Supplementary Information}
\newcommand{\beginsupplement}{%
        \setcounter{table}{0}
        \renewcommand{\thetable}{Supplementary \arabic{table}}%
        \setcounter{equation}{0}
       \renewcommand{\theequation}{S\arabic{equation}}%
        
        \setcounter{figure}{0}
        \renewcommand{\figurename}{\textbf{Supplementary Figure}}
}

\beginsupplement

\subsection*{Supplementary Note 1. Dot Potential}
We calculate the dot potential assuming a grounded graphene layer, $V_\mathrm{Gr} = 0$. As seen in Figure 1(b) in the main text, since $-\mathrm{e}V_\mathrm{Gr} = \mu_\mathrm{Gr}+ \mathrm{e}\phi(0)$, we have $\phi(0) = -\mu_\mathrm{Gr}/\mathrm{e}$. The electric field between the graphene and the source is: 
\begin{equation}
E = (\mu_\mathrm{Gr}/\mathrm{e}-V_\mathrm{Src})/d_\mathrm{Src}   
\end{equation}
The electric potential at the dot location $d_\mathrm{Dot}$ obeys:
\begin{equation}
\mathrm{e}\phi(d_\mathrm{Dot}) = \mathrm{e}\phi(0)+\mathrm{e}Ed_\mathrm{Dot}=-\mu_\mathrm{Gr}+(\mu_\mathrm{Gr}-\mathrm{e}V_\mathrm{Src})(d_\mathrm{Dot}/d_\mathrm{Src})    
\end{equation}

The dot energy $\mu_\mathrm{Dot}$ is shifted by $\mu_0$ from the potential at $z = d_\mathrm{Dot}$: 
\begin{equation}\label{eq_mu_dot_supp}
\mu_\mathrm{Dot}=\mu_0+\mu_\mathrm{Gr}\left(\frac{d_\mathrm{Dot}}{d_\mathrm{Src}}-1\right)-\mathrm{e}V_\mathrm{Src}\frac{d_\mathrm{Dot}}{d_\mathrm{Src}}
\end{equation}

\subsection*{Supplementary Note 2. Extracting DOS from Slope}

From the resonant features in Fig. 1(c), we can extract the density of states, and quantum capacitance, using two equations. The first is the integral equation (Equation (6) in the main text) for the graphene chemical potential $\mu_\mathrm{Gr}$:

\begin{equation}\label{eq:integral_eq_supp}
\int_{0}^{\mu_\mathrm{Gr}}\rho(E)dE = n_0 + \frac{C_\mathrm{Gate}V_\mathrm{Gate}+C_\mathrm{Src}V_\mathrm{Src}}{\mathrm{e}}-\frac{\mu_\mathrm{Gr}}{\mathrm{e^2}}(C_\mathrm{Gate}+C_\mathrm{Src})
\end{equation}
which relates $V_\mathrm{Gate}$, $V_\mathrm{Src}$ and $\mu_\mathrm{Gr}$. The solution to this equation defines the charge density $n_\mathrm{Gr}=\int_{0}^{\mu_\mathrm{Gr}}\rho(E)dE$ for any gate and bias voltage.

Along source onset line, where the dot is resonant with the source electrode, we have $\mu_\mathrm{Dot}=-\mathrm{e}V_\mathrm{Src}$. 
Plugging it into the source-dot resonance condition (Equation~\ref{eq_mu_dot_supp}) we have:
\begin{equation}
    -\mathrm{e}V_\mathrm{Src}=\mu_0+\mu_\mathrm{Gr}\left(\frac{d_\mathrm{Dot}}{d_\mathrm{Src}}-1\right)-\mathrm{e}V_\mathrm{Src}\frac{d_\mathrm{Dot}}{d_\mathrm{Src}}
\end{equation}
This yields the equation
\begin{equation}\label{eq:Vsrc_mugr_supp}
    V_\mathrm{Src}=\frac{\mu_\mathrm{Gr}}{\mathrm{e}}+\frac{\mu_0}{\mathrm{e}}\frac{d_\mathrm{Src}}{d_\mathrm{Dot}-d_\mathrm{Src}}
\end{equation}
Where the last equation means that to stay on resonance, $V_\mathrm{Src}$ will trace the graphene chemical potential up to a constant.
We use two equations: The integral equation, Equation~\ref{eq:integral_eq_supp}, and Equation~\ref{eq:Vsrc_mugr_supp}. These two depend on $V_\mathrm{Src}$, $V_\mathrm{Gate}$ and $\mu_\mathrm{Gr}$. We are interested in an expression for a slope in the ($V_\mathrm{Gate}$,\vsrc) plane: $\mathrm{d}V_\mathrm{Src} / \mathrm{d}V_\mathrm{Gate}$. To reach this, we plug Equation~\ref{eq:Vsrc_mugr_supp} into Equation~\ref{eq:integral_eq_supp}:


\begin{equation}
    \int_{0}^{\mu_\mathrm{Gr}}\rho(E)dE = n_0 -\frac{\mu_\mathrm{Gr}}{\mathrm{e^2}}C_\mathrm{Gate}
    +\frac{1}{\mathrm{e}}C_\mathrm{Gate}V_\mathrm{Gate}+\frac{1}{\mathrm{e}}C_\mathrm{Src}\left(\frac{\mu_0}{\mathrm{e}}\frac{d_\mathrm{Src}}{d_\mathrm{Dot}-d_\mathrm{Src}}\right)
\end{equation}

From which we isolate $V_\mathrm{Gate}$
\begin{equation}
    V_\mathrm{Gate}=-\frac{\mathrm{e}}{C_\mathrm{Gate}}n_0+\frac{\mu_\mathrm{Gr}}{\mathrm{e}}
    +\frac{\mathrm{e}}{C_\mathrm{Gate}}\int_{0}^{\mu_\mathrm{Gr}}\rho(E)dE  
    -\frac{C_\mathrm{Src}}{C_\mathrm{Gate}}\left(\frac{\mu_0}{\mathrm{e}}\frac{d_\mathrm{Src}}{d_\mathrm{Dot}-d_\mathrm{Src}}\right)
\end{equation}

Yielding
\begin{equation}
    \frac{\mathrm{d}V_\mathrm{Gate}}{\mathrm{d}\mu_\mathrm{Gr}}=\frac{1}{\mathrm{e}}+\frac{\mathrm{e}}{C_\mathrm{Gate}}\rho(\mu_\mathrm{Gr})
\end{equation}

From Equation~\ref{eq:Vsrc_mugr_supp}:
\begin{equation}
    \frac{\mathrm{d}V_\mathrm{Src}}{\mathrm{d}\mu_\mathrm{Gr}}=\frac{1}{\mathrm{e}}
\end{equation}

Thus
\begin{equation}
    \frac{\mathrm{d}V_\mathrm{Src}}{\mathrm{d}V_\mathrm{Gate}}=\frac{1}{1+\frac{\mathrm{e^2}}{C_\mathrm{Gate}}\rho(\mu_\mathrm{Gr})}
\end{equation}

\subsection*{Supplementary Note 3. Resonant Trajectory}

In the specific case where the quantum dot traces the spectrum of a Landau level, as in Figure 2, we can reach a special case of Equation (2) (Equation~\ref{eq:Vsrc_mugr_supp}), showing that $\Delta\mu_\mathrm{Gr} = \mathrm{e}\Delta V_\mathrm{Src}$ upon a transition between two Landau levels.
We calculate the shift in \vsrc\ between the two compressible plateaus corresponding to the difference between the $N$ and $N+1$ Landau levels. This is elucidated in the three schematic energy diagrams presented in Figure 2(a) in the main text, where the trajectory from (4,0) to (4,1) is broken into two stages.
First, is a transition from (i) to (ii) along a fixed \vsrc\ line, graphene shifts from ground state $N_\mathrm{G}$, with chemical potential $\mu_\mathrm{Gr}^N$ to ground state $N_\mathrm{G}+1$ with chemical potential $\mu_\mathrm{Gr}^{N+1}$. 
From Equation (1) (Equation~\ref{eq_mu_dot_supp})  we find the dot energy shift is:
\begin{equation}
\mathrm{\Delta\mu}_\mathrm{Dot}^{\mathrm{(i)-(ii)}}=\mathrm{\Delta \mu}_\mathrm{Gr}(d_\mathrm{Dot}/d_\mathrm{Src}-1)
\end{equation}
Here $\mathrm{\Delta \mu_\mathrm{Gr}} = \mu_\mathrm{Gr}^{N+1}-\mu_\mathrm{Gr}^N$ is the difference in the graphene chemical potential upon increase of a single Landau level.
In the second leg of the trajectory, from (ii) to (iii), the dot is brought back to resonance with the same $N_\mathrm{I}$ through a shift in \vsrc\ while keeping $n_\mathrm{Gr}$ (and $\mu_\mathrm{Gr}$) fixed. Here Equation (1) yields
\begin{equation}
    \Delta\mu_\mathrm{Dot}^{(ii)-(iii)} = \mathrm{e}\Delta V_\mathrm{Src}(-d_\mathrm{Dot}/d_\mathrm{Src})
\end{equation}
The total shift in $\mu_\mathrm{Dot}$ from (i) to (iii) is
\begin{equation}
\mathrm{\Delta\mu_\mathrm{Dot}^{(i)-(iii)}} = \mathrm{\Delta\mu}_\mathrm{Gr}(d_\mathrm{Dot}/d_\mathrm{Src}-1)+\mathrm{e}\Delta V_\mathrm{Src}(-d_\mathrm{Dot}/d_\mathrm{Src})    
\end{equation}

Demanding that the dot remains resonant with the same Landau level requires:
\begin{equation*}
\mathrm{\Delta\mu}_\mathrm{Dot}^{\mathrm{(i)-(iii)}}=\mathrm{\Delta\mu}_\mathrm{Gr}    
\end{equation*}
we have
\begin{equation*}
    \mathrm{\Delta\mu}_\mathrm{Gr}=\mathrm{\Delta\mu}_\mathrm{Gr}(d_\mathrm{Dot}/d_\mathrm{Src}-1)+\mathrm{e}\mathrm{\Delta V}_\mathrm{Src}(-d_\mathrm{Dot}/d_\mathrm{Src})
\end{equation*}
which yields
\begin{equation} 
\mathrm{\Delta\mu}_\mathrm{Gr} = \mathrm{e}\mathrm{\Delta V}_\mathrm{Src}
\end{equation}
This relation is general for trajectories where the dot is kept in resonance with a specific spectral feature. In our case it means that tuning the system from from (i) to (iii), while changing the graphene ground state from $N_\mathrm{G}$ to $N_\mathrm{G}+1$ results in $\mathrm{e}\mathrm{\Delta V}_\mathrm{Src} = \mu_\mathrm{Gr}^{N+1}-\mu_\mathrm{Gr}^N$. This is also demonstrated in panel (iii) of Figure 2 in the main text, where we find that $\phi(z)$ corresponding to state (i), plotted in red, is parallel to the new $\phi(z)$ corresponding to state (iii),  plotted in black.

\subsection*{Supplementary Note 4. Raw Data}

\begin{figure}[h]

\includegraphics[width=\linewidth]{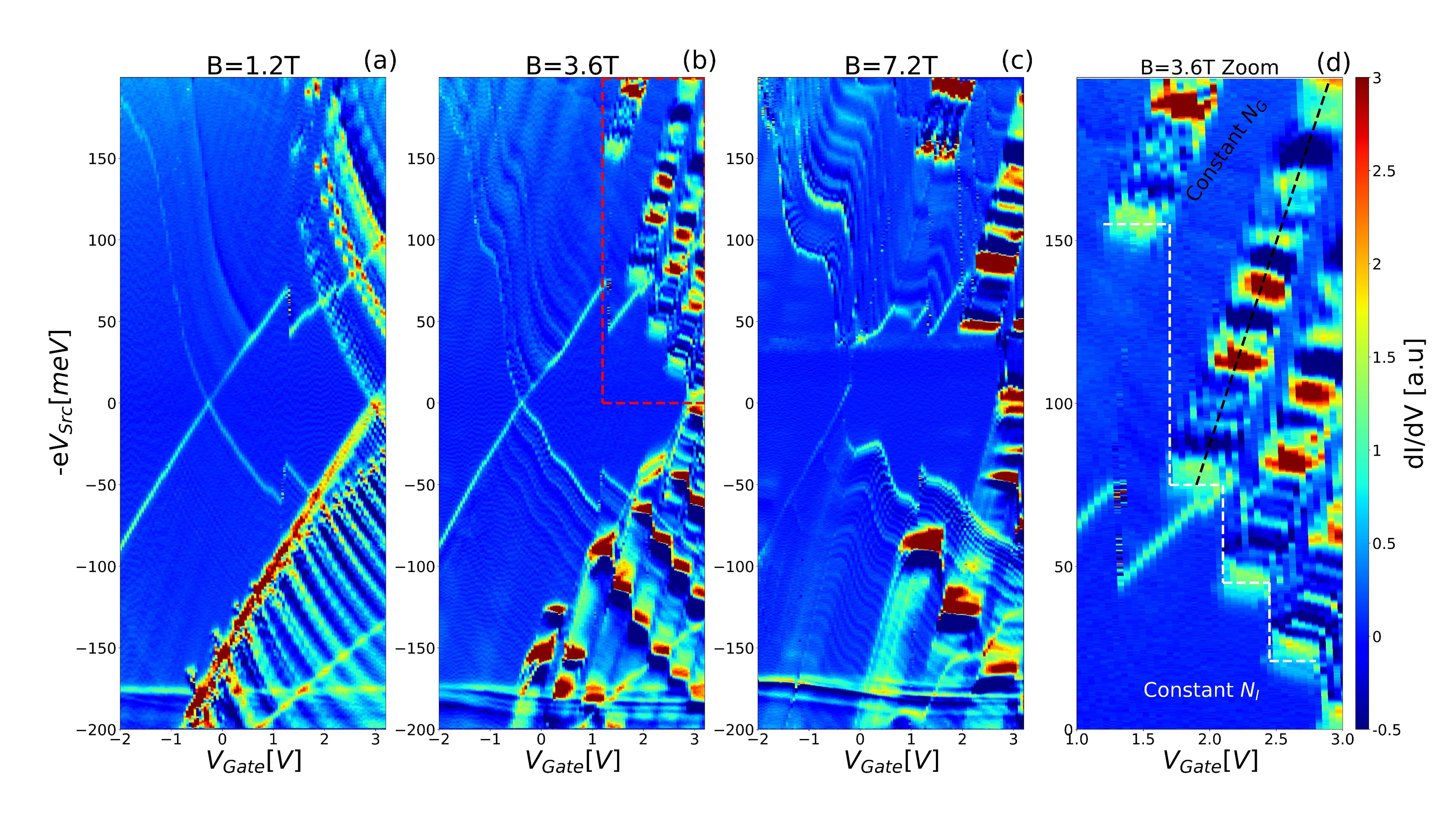}
    
    \caption{ Raw Data Finite Field Measurements 
    (\textbf{a-c})  $\mathrm{d} I/ \mathrm{d}V$ maps at $B=1.2$ T, $B=3.6$ T and $B=7.2$ T. The dashed square in (b) marks the zoomed in region for panel (d). (\textbf{d}) Zoom in on panel (b). Constant injection and constant ground state trajectories are marked (white and black respectively)}
 
    \label{fig:Raw_Field}

\end{figure}

In the main text, the horizontal axis used in Fig. 2 and Fig. 4 shows $\tilde{V}=V_\mathrm{Gate}+\frac{C_\mathrm{Src}}{C_\mathrm{Gate}}V_\mathrm{Src}$, defined such that the graphene equal density lines are vertical - leading to easier identification of the energy difference between Landau levels.
In Supplementary \Cref{fig:Raw_Field} we show the raw data, where the horizontal axis is \vg. Panels (a-c) show the same measurements presented in Fig. 2 while panel (d) shows the region marked by a red square in (b), zoomed in. This representation of the data enables appreciation of the sharpness of the transition between adjacent ground states.

From Supplementary \Cref{fig:Raw_Field}(d) it is evident that the transition, along a constant injection trajectory, between adjacent ground states is vertical. This embodies the fact that once a Landau level is full, further gating the graphene with little additional charge shifts its spectrum by the energy difference between levels. Along a constant injection trajectory, \vsrc\ compensates for this shift resulting in the vertical staircase structure.

\subsection*{Supplementary Note 5. Device Details}

\begin{figure*}[h]

\includegraphics[width=\linewidth,height=8cm]{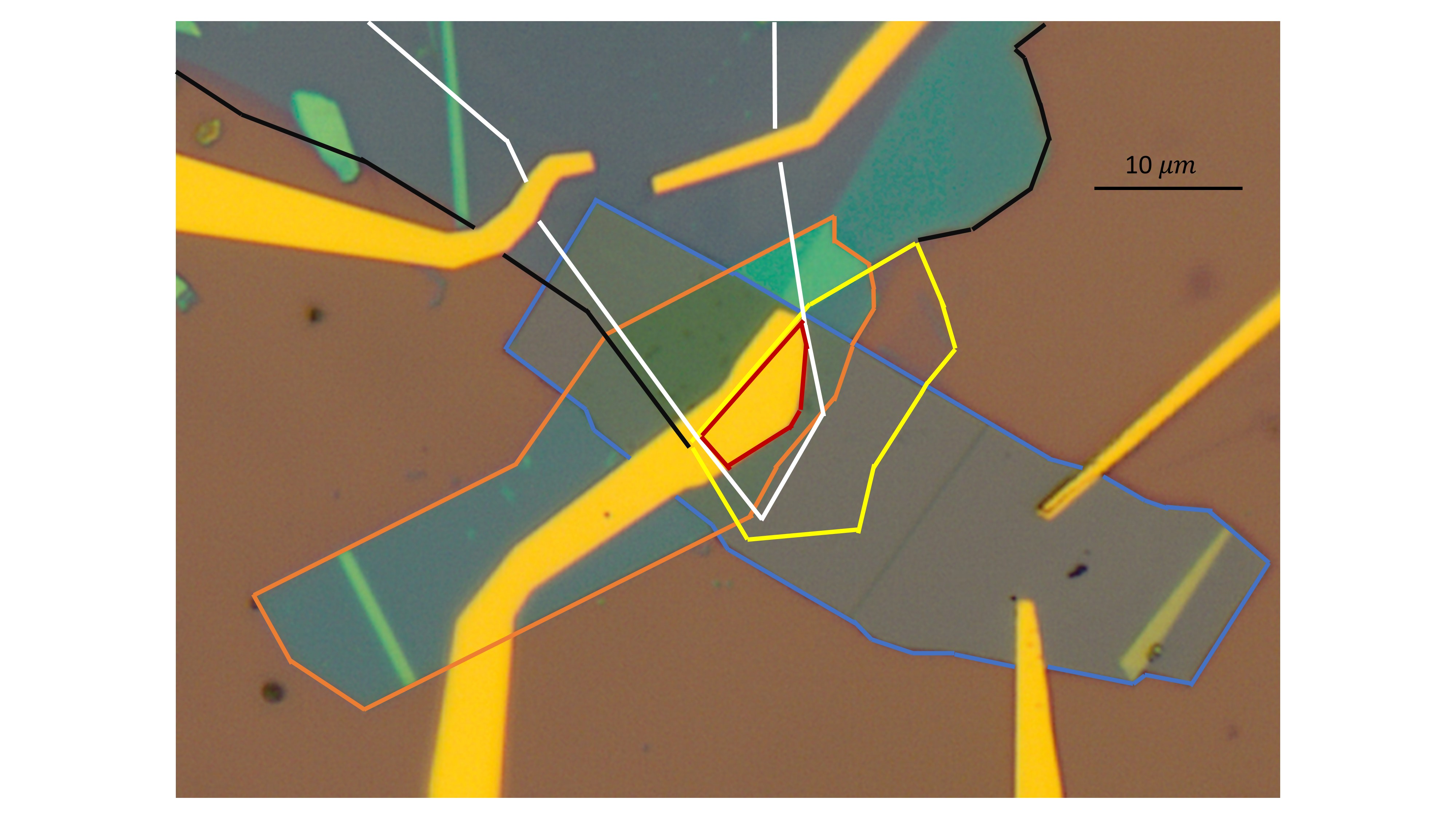}

    \caption{Optical microscope image of the device. The different layers are outlined from bottom to top - graphite (blue), barrier hBN (yellow), graphene (white) and a top dielectric hBN (orange). The barrier hBN flake is attached to a thicker region outlined in black. The gate electrode is patterned on top of the dielectric hBN. Gated junction area is approximately 45 $\mathrm{\mu m^2}$. The gated region is outlined in red.}

    \label{fig:Device}

\end{figure*}

The data in this work originates from a single device consisting of two dots within a region of approximately 45 $ \mathrm{\mu m^2} $ pictured in Supplementary \Cref{fig:Device}.

In Supplementary \Cref{fig:Device} one can see an overlap between graphene (white), barrier hBN (yellow) and graphite (blue) away from the gated region (red). 
This leads to the possibilty of un-gated dots and might explain the constant feature near -180 meV in Supplementary \Cref{fig:Raw_Field} as an additional un-gated dot. The area of the un-gated region is approximately 30 $ \mathrm{\mu m^2}$. All together, our junction has 3 dots over the region of 75 $\mathrm{\mu m^2}$ which leads to the estimate of one dot per 25 $\mathrm{\mu m^2}$.  We note that this represents a smaller dot density than reported in Ref.~\cite{Greenaway2018}.
Based on these dimensions, and on dimensions of devices measured by others~\cite{Greenaway2018,Chandni2015,Chandni2016}, to obtain transport-active dots, one requires a gated junction area should be 20-30 $ \mathrm{\mu m^2}$ and barrier hBN thickness 3-6 layers.

\section*{References}
\def\bibsection{}

\end{document}